# Observation of a Long-Range Interaction between Semiconductor Quantum Dots

E.W. Bogaart, J.E.M. Haverkort, and R. Nötzel

*Eindhoven University of Technology, Department of Applied Physics*
*P.O. Box 513, 5600MB Eindhoven, The Netherlands*

**Abstract.** We demonstrate electromagnetic interaction between distant quantum dots (QDs), as is observed from transient pump-probe differential reflectivity measurements. The QD-exciton lifetime is measured as a function of the probe photon energy and shows a strong resonant behavior with respect to the QD density of states. The observed exciton lifetime spectrum reveals a subradiance-like coupling between the QD, with a 12 times enhancement of the lifetime at the center of the ground state transition. This effect is due to a mutual electromagnetic coupling between resonant QDs, which extends over distances considerably beyond the nearest neighbor QD-QD separation.

**Keywords:** quantum dots, time-resolved, cooperative phenomena, electromagnetic interaction.

**PACS:** 78.47.+p, 42.50.Fx, 78.67.Hc    (Proceedings 28th ICPS conference, July 2006, Vienna Austria)

The manipulation of the spontaneous lifetime of an emitter, e.g., an excited quantum dot (QD), is of particular interest for fundamental physics as well as for future applications in quantum logic devices [1]. Modification of the local density of electromagnetic states in the vicinity of the emitter, known as the Purcell effect, by inserting the emitter into a nanocavity [2,3] allows for inhibition and enhancement of the spontaneous emission lifetime.

Optical excitation of a QD-exciton has a profound impact on the QD permittivity near its resonance. The permittivity of an excited QD can be modeled [4] with a Lorentzian contribution and has a calculated peak value of $\varepsilon_{QD} \sim 10^6$ at resonance [4,5], which is much larger than the permittivity of the GaAs host medium ($\varepsilon_h = 11.5$). Thus, excited QDs form a high contrast permittivity landscape providing strong dipole scattering of the electromagnetic fields [6]. In other words, an excited QD strongly modifies the local polarization for resonant light, which is likely to couple with other excited QDs. Hence, the initially isolated QDs are collectively coupled by the electromagnetic field.

We demonstrate a long-range electromagnetic interaction between spectrally identical QDs. From time-resolved differential reflection data, in which the QD reflectance is monitored as a function of the probe photon energy [7], we observe a significant enhancement of the exciton lifetime in a semiconductor QD ensemble without any external manipulation of the electromagnetic field. This collective effect is comparable in magnitude with the lifetime enhancement reported for a single QD-exciton within an artificial nanocavity [2,8]. Our observations indicate that the QDs mutually couple over distances considerably beyond the nearest neighbor separation.

The experiments are performed on self-assembled InAs/GaAs QDs grown by molecular beam epitaxy on a (100) GaAs substrate. The nanostructure studied here, is the same sample as reported in Ref. 7. Atomic force microscopy images of uncapped QDs reveal the formation of QDs with a density of $2.8 \times 10^{10}$ cm$^{-2}$.

The QD-exciton lifetime is investigated by means of pump-probe time-resolved differential reflection spectroscopy [7]. The QDs are non-resonantly excited using short laser pulses from a 2 ps mode-locked Ti:sapphire laser, creating free carriers which are subsequently captured in the QDs. The carrier-induced differential reflection $\Delta R/R_0$ is monitored by tuning the probe laser into resonance with the transition energy of the QD ensemble [7]. The 2 ps probe pulses are generated from an optical parametric oscillator, synchronically pumped by the Ti:sapphire laser.

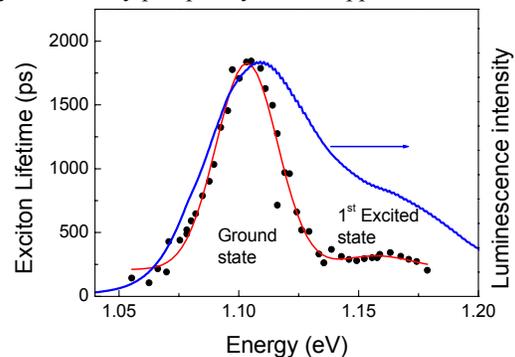

**Figure 1.** Exciton lifetime and photoluminescence intensity versus the exciton transition energy (300 W/cm$^2$, 5 K). The resonant-like behavior is due to electromagnetic coupling of energetically identical QDs.

The main consequence of the electromagnetic coupling due to the collective polarizability of the structure [6,9] is a profound modification of the exciton lifetime, as is depicted by the spectrum in Fig. 1. The exciton lifetime spectrum shows a pronounced resonant-like behavior with a 12 times enhancement near the center of the ground state energy distribution at 1.107 eV. Surprisingly, also for the QD first excited state at 1.163 eV an enhancement of the exciton lifetime is observed.

Another pronounced effect is that the exciton lifetime spectrum is considerably narrower than the photoluminescence (PL) spectrum. Analysis of the lifetime spectrum using Gaussian fits reveals a spectral width of the ground (excited) state of 27 meV (23 meV) while the spectral width of the PL is 44 meV.

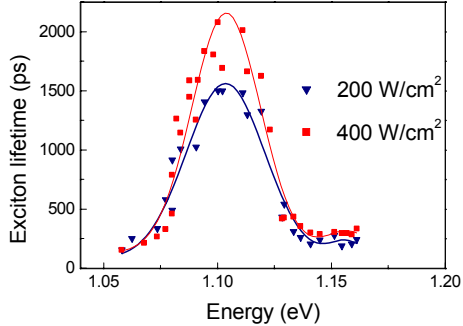

**Figure 2.** *Excitation density dependence of the exciton lifetime at 200 and 400 W/cm$^2$, obtained at 5 K.*

Figure 2 shows additional evidence for an electromagnetic coupling between optically excited QDs. The exciton lifetime exhibits a strong dependence on the pump excitation density. We observe a maximum exciton lifetime of 1.5 ns at 200 W/cm$^2$ pump excitation which increases to 2.1 ns at 400 W/cm$^2$ pump excitation. The strong dependence of the exciton lifetime on the pump power indicates that the exciton lifetime is governed by the fraction of excited QDs within the sample. Thus, the strength of the electromagnetic coupling can be optically tuned. The experimental results as depicted in Figs. 1 and 2 are clearly in contradiction with the conventional hypothesis of a weak electromagnetic coupling of QDs.

In order to explain and to analyze the observed lifetime enhancement, we use and significantly extend the electromagnetic response theory for a periodic array of identical and spherical QDs [6] with an effective exciton volume $V_{QD}$. Calculations show that the exciton emission rate $\Gamma^{coupled}$ of a QD-array - with transition energy $\hbar\omega_0$ - can be written as

$$\Gamma^{coupled} = \Gamma^{isolated} - \frac{k_0\sqrt{\varepsilon_h}\omega_{QD}V_{QD}}{2d_{QD}^2}. \quad (1)$$

Here $\Gamma^{isolated}$ is a weighted sum of the dephasing rate [10] and the radiative emission rate of a single isolated QD. $k_0$ and $d_{QD}$ denote the vacuum wave vector and the spacing of the QDs within the two-dimensional periodic lattice, respectively, and $\omega_{QD}$ is the phenomenological parameter proportional to the QD oscillator strength [6]. Equation (1) shows that the lifetime $1/\Gamma^{coupled}$ is governed by the lattice spacing $d_{QD}$.

The resonant and optically excited QDs form a sub-ensemble which can be approximated by an ordered array [11] with an average lattice spacing $d^{res}_{QD}$. This means that the whole QD ensemble is divided into smaller sub-ensembles each with their own sub-ensemble density $N(\omega)$. The average distance between the QDs within each sub-ensemble depends on the location of these QDs within the overall size distribution, and is governed by $1/d^{res}_{QD} = N(\omega) = \sqrt{N_{QD}\,G(\omega)}$. $N_{QD}$ denotes the area QD density and $G(\omega)$ is a Gaussian line shape function representing the inhomogeneously broadened density of states of the QD ensemble. Hereby Eq. (1) is written as

$$\Gamma^{coupled}(\omega) = \Gamma^{isolated}(\omega) - 2\frac{k_0\sqrt{\varepsilon_h}\omega_{QD}V_{QD}}{2}N_{QD}G^2(\omega). \quad (2)$$

Equation (2) proves the observed functional behavior, as observed from Figs. 1 and 2. The additional factor two in the second term on the righthand side of Eq. (2) takes into account the QD spin degeneracy. Equation (2) also predicts that the width of the exciton lifetime spectrum is a factor $\sqrt{2}$ narrower than the QD distribution $G(\omega)$ due to the quadratic dependence. This prediction is indeed observed experimentally (see Fig. 1)

Applying the electromagnetic response theory to our QDs, our observed 12 times enhancement of the exciton lifetime corresponds to an average distance of $d^{res}_{QD} = 490 \pm 20$ nm between the radiatively coupled QDs at the center of the QD-size distribution, as is illustrated in Fig. 1. We note that the wings of the QD-exciton lifetime spectrum are thus governed by an even larger separation between the mutually interacting QDs.